\documentclass[aps,prd,onecolumn,groupedaddress,showpacs,showkeys, nofootinbib,amssymb]{revtex4}
\usepackage[T1]{fontenc}
\usepackage[latin1]{inputenc}
\usepackage{graphicx}
\usepackage[english]{babel}
\usepackage{bm}
\usepackage{amsmath}
\usepackage{amssymb}
\usepackage{amsfonts}
\usepackage{epsfig}
\usepackage{colordvi}
\usepackage{color}

\begin{document}

\title{Noether Symmetry Approach in Gauss-Bonnet Cosmology}
\author{Salvatore Capozziello$^{1,2,3}$, Mariafelicia De Laurentis$^{1,2,4}$\footnote{e-mail address: mfdelaurentis@tspu.edu.ru}, Sergei D. Odintsov$^{4,5,6}$}
\affiliation{$^{1,2}$Dipartimento di Fisica, Università
di Napoli {}``Federico II'', Compl. Univ. di
Monte S. Angelo, Edificio G, Via Cinthia, I-80126, Napoli, Italy\\
INFN Sezione  di Napoli, Compl. Univ. di
Monte S. Angelo, Edificio G, Via Cinthia, I-80126, Napoli, Italy.}
\affiliation{$^3$Gran Sasso Science Institute (INFN), Via F. Crispi 7, I-67100, L' Aquila, Italy.}
\affiliation{$^{4}$Tomsk State Pedagogical University, 634061 Tomsk and National Research Tomsk State University, 634050 Tomsk, Russia}
\affiliation{$^{5}$Instituci\'o Catalana de Recerca i Estudis Avancats (ICREA), Barcelona, Spain} 
\affiliation{$^{6}$  Consejo Superior de Investigaciones Cientificas,
ICE/CSIC-IEEC, Campus UAB, Torre C5-Parell-2a pl, E-08193 Bellaterra (Barcelona), Spain }
\date{\today}

\begin{abstract}
We discuss the Noether Symmetry Approach in the framework of Gauss-Bonnet
 cosmology showing that the functional form of the $F(R, {\cal G})$ function, where $R$ is the Ricci scalar and ${\cal G}$ is the Gauss-Bonnet topological invariant,  can be determined by the presence of symmetries. Besides, the method allows to find out exact  solutions due to the reduction of cosmological dynamical system and the presence of conserved quantities. Some  specific cosmological models are worked out.  
 \end{abstract}

\maketitle

\section{Introduction}
\label{uno}
The investigation of alternative theories of gravity mainly arises from  cosmology, quantum field theory and some open issues in astrophysics. Problems like initial singularity, horizon and flatness point out that the Standard Cosmological Model,  based on Particle Standard Model and General Relativity, fails when one wants to describe the whole universe, especially at extreme regimes of ultraviolet scales  \cite{guth,weinberg}. In addition, General Relativity does not work as a fundamental theory capable of giving a quantum description of space-time. The results of these shortcomings and, first of all,   the absence of a final  Quantum Theory of Gravity  claim for modifications of General Relativity that have been introduced in order to approach, at least, a  semiclassical description  towards   quantization. These theories are aimed to address the problem of gravitational interaction   by adding, for example,   non-minimally coupled scalar fields or higher-order curvature invariants into the Hilbert-Einstein and then are often called Extended Theories of Gravity \cite{PhysRepnostro,OdintsovPR,5,6,Mauro,faraoni,9,10,libri,libroSV,libroSF}.
In particular, if we want to obtain the effective action of quantum gravity on scales closer to the Planck length,  we need to introduce these corrective terms \cite{vilkovisky, Buchbinder}.
Furthermore, in order to construct a renormalizable theory of gravity, one can include a wide number of higher-order terms  of curvature invariants in the Lagrangian, such as $R^2$ $R_{\mu\nu}R^{\mu\nu}, R_{\mu\lambda\nu\sigma}R^{\mu\lambda\nu\sigma}$ among others, considering also the possibility to include   the Gauss-Bonnet topological invariant. Specifically,  several authors have constructed  renormalizable  gravity models by introducing quadratic terms in curvature invariants giving for example a detalied analysis of the one-loop divergences and schemes of quantization  \cite{stelle,stelle1,christensen,barth}. For a general review see \cite{Buchbinder}.
It is well know that actions involving    a finite number of  power laws of curvature corrections  and their derivatives can be considered as   low-energy approximations to some fundamental theory of gravity like strings or supergravity \cite{fradkin,IG}.
In particular, if we consider Lagrangians with higher-order terms or arbitrary derivatives in curvature invariants, it is possible to obtain  non-local Lagrangians that give rise to some characteristic length of the order of Planck length \cite{birrel}.
Furthermore, in order to obtain  ghost-free actions from string/M-theory, we need to introduce quadratic curvature corrections to the Einstein Hilbert action proportional to the Gauss-Bonnet term \cite{OdintsovPR,cognola,odi06}. 

Recently, new generalized versions of Gauss-Bonnet gravity have been  considered.  The action is given by  general functions of the Ricci scalar $R$ and  the Gauss-Bonnet  topological invariant ${\cal G}$,  that is $f(R,{\cal G})$ \cite{F(GR)-gravity,17,18,19,20,21,22,23,24,25,26,27,28,29,30,31,32,ester,granda1,granda2}. These theories are  stable and  capable of describing   the present acceleration of the universe as well as the phantom behavior, the quintessence behavior  and the transition from acceleration to deceleration phases.
In this sense, they are effective theories working also at infrared scales.  In principle, this kind of Extended Theories of Gravity can reproduce  models able to mimic the $\Lambda$CDM model, as well as other  cosmological solutions and suitable perturbation schemes  within different standard scenarios \cite{diego,myrz,diego3,Felice_malo}. Also, the Parametrized Post-Newtonian expansion of generalized Gauss-Bonnet models has been   worked out  \cite{antonio} as well as  spherically  symmetric solutions \cite{zerb}.

In this paper, we are going to discuss if it is possible to fix the form of   $f(R,{\cal G})$-Lagrangians by the existence of symmetries. In particular we are using the so called Noether Symmetry Approach \cite{cimento,hamilton,greci,pla,pla2,marek,india,stabile,stabile1}.
 We will show that,  asking for the existence of 
Noether symmetries, it is to possible to select physically interesting forms of $f(R,{\cal G})$ and
the existence of  symmetries allows to select constants of
motion  that reduce dynamics. Furthermore,
reduced dynamics results exactly solvable by a straightforward change of
variables where a cyclic coordinate is present.
As we will see, the method is twofold: from one side, the existence of symmetries allows to solve exactly the dynamics; from the other side, the Noether charge can always be related to some observable quantity.
The general philosophy is that both $R$ and ${\cal G}$ behave like effective scalar fields as soon as suitable Lagrange multipliers are introduced into dynamics \cite{makarenko}. This fact allows to define a suitable configuration space ${\cal Q}\equiv \{a,R,{\cal G}\}$, where $a$ is the Friedman-Robertson-Walker scale factor: one can search for the invariance of a Lie vector field $X$ by the Lie derivative $L_{X}$ acting on the point-like Lagrangian ${\cal L}({\dot a}, a, {\dot R}, R, {\dot{\cal G}},{\cal G})$, that is $L_X {\cal L} =0$.

 The layout of the paper is the following. In Sec. \ref{due} we sketch the main ingredients of  $f(R,{\cal G})$. Sec. \ref{tre} is devoted to the derivation of the Friedman-Robertson-Walker  cosmological equations starting from the point-like Lagrangian. The Noether Symmetry Approach is discussed  in detail  in Sec. \ref{quattro} and some cosmological models, related to the existence of the symmetries, are worked out in Sec. \ref{cinque}.
Conclusions are drawn in Sec. \ref{sei}.
\section{Gauss-Bonnet  gravity}
\label{due}

Let us start from  the most general action for modified Gauss-Bonnet gravity in 4-dimensions. It is  
\begin{equation}
{\cal A}=\int d^{4}x\sqrt{-g}\left[\frac{1}{2\kappa^2}F(R,{\cal G})+\mathcal{L}_{m}\right]\,,
\label{action1}
\end{equation}
where $\kappa^2=8\pi G_N$, $G_N$ is the Newton constant and $\mathcal{L}_m$ is the standard matter Lagrangian and  $g$  the determinant of the metric. This Lagrangian is constructed by considering only the metric tensor and no extra vector or spin degree of freedom is considered.
The symbol ${\cal G}$ indicates the Gauss-Bonnet invariant
\begin{equation}
 {\cal G}\,\equiv\,R^2-4R_{\mu\nu}R^{\mu\nu}+R_{\mu\nu\lambda\sigma}R^{\mu\nu\lambda\sigma}\,,
\label{GB}
\end{equation} 
that is a combination of the Riemann tensor $R_{\mu\nu\lambda\sigma}$, the Ricci tensor $R_{\mu\nu}$ and the Ricci scalar $R=g^{\mu\nu}R_{\mu\nu}$. It is important to remember that, in 4-dimension, any linear combination of the Gauss-Bonnet invariant does not contribute to the effective
Lagrangian. Furthermore,  in 4-dimensions, we have only two non-zero Lovelock scalars \cite{Felice_malo,kazu,sebastiani}.Then, by varying expression in Eq. (\ref{action1}) with respect to the metric tensor $g_{\mu\nu}$, the modified Einstein field equations are obtained \cite{antonio,cognola},

\begin{eqnarray}
0\,&=&\,\kappa^{2} T^{\mu\nu}+\frac{1}{2}g^{\mu\nu}F(R, {\cal G})-2 \frac{\partial F(R,{\cal G})}{\partial {\cal G}}RR^{\mu\nu}+4 \frac{\partial F(R,{\cal G})}{\partial {\cal G}}R^{\mu}_{\rho}R^{\nu\rho}-2\frac{\partial F(R,{\cal G})}{\partial {\cal G}}R^{\mu\rho\sigma\tau}R^{\nu}_{\rho\sigma\tau}+\nonumber\\&&\nonumber\\&&-\,4 \frac{\partial F(R,{\cal G})}{\partial {\cal G}}R^{\mu\rho\sigma\nu}R_{\rho\sigma}+2\left(\nabla^{\mu}\nabla^{\nu} \frac{\partial F(R,{\cal G})}{\partial {\cal G}}\right)R
-\,2g^{\mu\nu}\left(\nabla^{2} \frac{\partial F(R,{\cal G})}{\partial {\cal G}}\right)R-\,4\left(\nabla_{\rho}\nabla^{\mu} \frac{\partial F(R,{\cal G})}{\partial {\cal G}}\right)R^{\nu\rho}+\nonumber\\&&\nonumber\\&&-\,4\left(\nabla_{\rho}\nabla^{\nu} \frac{\partial F(R,{\cal G})}{\partial {\cal G}}\right)R^{\mu\rho}+4\left(\nabla^{2} \frac{\partial F(R,{\cal G})}{\partial {\cal G}}\right)R^{\mu\nu}+4g^{\mu\nu}\left(\nabla_{\rho}\nabla_{\sigma} \frac{\partial F(R,{\cal G})}{\partial {\cal G}}\right)R^{\rho\sigma}+\nonumber\\&&\nonumber\\&&
-\,4\left(\nabla_{\rho}\nabla_{\sigma} \frac{\partial F(R,{\cal G})}{\partial {\cal G}}\right)R^{\mu\rho\nu\sigma}- \frac{\partial F(R,{\cal G})}{\partial {\cal G}}R^{\mu\nu}+\nabla^{\mu}\nabla^{\nu} \frac{\partial F(R,{\cal G})}{\partial R}-g^{\mu\nu}\nabla^{2}\frac{\partial F(R,{\cal G})}{\partial R}\,,
\label{FE}
\end{eqnarray}
where $\nabla$ is for the  covariant derivative and 
${\displaystyle T_{\mu\nu}=\frac{-2}{\sqrt{-g}}\frac{\delta(\sqrt{-g}\mathcal{L}_m)}{\delta
g^{\mu\nu}}}$ the energy momentum tensor. Let us note that  General Relativity is immediately recovered for  $F(R,{\cal G})=R$.

\section{Gauss-Bonnet cosmology}
\label{tre}

\subsection{Reducing to a canonical point-like Lagrangian }
\label{tre1}

In order to calculate the cosmological equations, we need to deduce a point-like canonical Lagrangian ${\cal L}(a,{\dot a}, R,{\dot R},{\cal G},{\dot {\cal G}})$ from the action (\ref{action1}) where ${\cal Q}\equiv\{a,R,{\cal G}\}$ is the configuration space
and ${\cal TQ}\equiv \{a,\dot{a}, R, \dot{R}, {\cal G},{\dot {\cal G}}\}$ is the corresponding tangent
space  on which ${\cal L}$ is defined.  The variable $a(t)$,
$R(t)$ and ${\cal G}(t)$ are the scale factor, the  Ricci scalar and the Gauss-Bonnet invariant defined in the Friedman-Roberston-Walker
metric, respectively. 
Its Euler-Lagrange equations are

\begin{eqnarray}
\frac{d}{dt}\frac{\partial {\cal L}}{\partial {\dot a}}=\frac{\partial {\cal L}}{\partial  a}\,, \qquad
\frac{d}{dt}\frac{\partial {\cal L}}{\partial {\dot R}}=\frac{\partial {\cal L}}{\partial  R}\,,\qquad
\frac{d}{dt}\frac{\partial {\cal L}}{\partial {\dot {\cal G}}}=\frac{\partial {\cal L}}{\partial  {\cal G}}\,,
\label{moto3}
\end{eqnarray}

with the energy equation
\begin{eqnarray}
E_{\cal L}= \frac{\partial {\cal L}}{\partial {\dot a}}{\dot a}+\frac{\partial {\cal L}}{\partial {\dot R}} {\dot R}+\frac{\partial {\cal L}}{\partial {\dot {\cal G}}} {\dot {\cal G}}-{\cal L}=0\,.
\label{energy}
\end{eqnarray}
Here the dot indicates the derivatives with respect to the cosmic time $t$.
One can use the method of the Lagrange
multipliers to set  $R$  and ${\cal G}$ as  constraints for  dynamics. Selecting
the suitable Lagrange multiplier and integrating by parts to eliminate higher than one time derivatives, the
Lagrangian ${\cal L}$ becomes canonical. Using physical units $G_N=c=\hbar=1$ and the signature $(+,-,-,-,)$, we have

\begin{eqnarray}
 {\cal A}=2\pi^2\int dt\,a^3 \left\{  F(R,{\cal G})-\lambda_1 \left[R+6\left( \frac{{\ddot a}}{a}+\frac{{\dot a}^2}{a^2}\right)\right]-\lambda_2\left[{\cal G}-24\left( \frac{{\ddot a}{\dot a}^2}{a^3}\right)\right]     \right\}\,.
\end{eqnarray}
Here the definitions of the Ricci scalar and the Gauss-Bonnet invariant in Friedman-Robertson-Walker metric have been adopted, that is  

\begin{eqnarray}
\label{consistency}
R=-6\left( \frac{{\ddot a}}{a}+\frac{{\dot a}^2}{a^2}\right)\,, \qquad {\cal G}= 24\left(\frac{{\ddot a}{\dot a}^2}{a^3}\right)\,.
\end{eqnarray}
A spatially flat Friedman-Roberson-Walker spacetime has been considered for the sake of simplicity. The Lagrange multipliers  $\lambda_{1,2}$ are given by varying the action with respect to $R$ and ${\cal G}$, that is

\begin{eqnarray}
\lambda_1= \frac{\partial F(R,{\cal G})}{\partial  R}\,,\qquad \lambda_2= \frac{\partial F(R,{\cal G})}{\partial {\cal G}}\,,
\end{eqnarray}
then the above action becomes
\begin{eqnarray}
 {\cal A}=2\pi^2\int dt \left\{ a^3  F(R,{\cal G})- a^3  \frac{\partial F(R,{\cal G})}{\partial  R} \left[R+6\left( \frac{{\ddot a}}{a}+\frac{{\dot a}^2}{a^2}\right)\right]-a^3\frac{\partial F(R,{\cal G})}{\partial {\cal G}}\left[{\cal G}-24 \left(\frac{{\ddot a}{\dot a}^2}{a^3}\right)\right] \right\}\,.
\end{eqnarray}
After an integration by parts, the point-like Lagrangian assumes the following form

\begin{eqnarray}
 {\cal L}= 6 a {\dot a}^2  \frac{\partial F(R,{\cal G})}{\partial  R}+ 6 a^2  {\dot a} \frac{d}{dt} \left( \frac{\partial F(R,{\cal G})}{\partial  R}\right)-8  {\dot a}^3  \frac{d}{dt}  \left(\frac{\partial F(R,{\cal G})}{\partial  {\cal G}}\right)    +a^3\left[ F(R,{\cal G})-R\, \frac{\partial F(R,{\cal G})}{\partial  R}-{\cal G} \frac{\partial F(R,{\cal G})}{\partial {\cal G}}\right]\,,\label{PointLagra}
\end{eqnarray}
which is a canonical function of 3 coupled fields $a$, $R$ and ${\cal G}$  depending on time $t$.

It is important to stress that the Lagrange multipliers have been opportunely chosen by considering the definition of the Ricci curvature scalar $R$ and the 
Gauss-Bonnet invariant ${\cal G}$. This fact allows us to consider the constrained dynamics as canonical. 

It is interesting to consider some important  cases of the Lagrangian (\ref{PointLagra}). For  $F(R,{\cal G})=R$,  the  Lagrangian  of General Relativity is recovered. In this case we have
\begin{eqnarray}
 {\cal L}=6a{\dot a}^2 +a^3R\,,
\end{eqnarray}
that, after developing $R$, easily reduces to ${\cal L}=-3a{\dot a}^2$, the standard point-like Lagrangian of Friedman-Robertson-Walker cosmology. 

In the case $F(R,{\cal G})=f(R)$, we have the standard $f(R)$-gravity \cite{PhysRepnostro}
\begin{eqnarray}
 {\cal L}= 6 a {\dot a}^2 f'(R)+ 6 a^2  {\dot a}{\dot R}f''(R)+a^3\left[f(R)-Rf'(R)\right]\,,\label{LfR}
\end{eqnarray}
while a pure Gauss-Bonnet cosmology is recovered for  $F(R,{\cal G})=f({\cal G})$, and then
\begin{eqnarray}
 {\cal L}= 8 {\dot a}^3 {\dot {\cal G}}f''({\cal G})+a^3\left[f({\cal G})-{\cal G}f'({\cal G})\right]
 \label{LfG}\,.
\end{eqnarray}
Clearly,  these cases deserve a specific investigation.

\subsection{The cosmological  equations}
\label{tre2}

 Let us now derive the Euler-Lagrange equations from Eqs. (\ref{moto3} - \ref{energy}).  They can be also deduced from the fields Eqs.(\ref{FE}). They are 
 
 \begin{eqnarray} 
\left[ \left(\frac{\dot a}{a}\right)^2+2\frac{{\ddot a}}{a}\right]\frac{\partial F(R,{\cal G})}{\partial  R}&+& \frac{d^2}{dt^2} \left(\frac{\partial F(R,{\cal G})}{\partial  R}\right)+2\frac{\dot a}{a} \frac{d}{dt} \left(\frac{\partial F(R,{\cal G})}{\partial  R}\right)-8 \frac{{\dot a}{\ddot a}}{a^2}  \frac{d}{dt} \left(\frac{\partial F(R,{\cal G})}{\partial  {\cal G}}\right) \nonumber\\&&\nonumber\\&&- 4  \left(\frac{\dot a}{a}\right)^2 \frac{d^2}{dt^2}  \left(\frac{\partial F(R,{\cal G})}{\partial  {\cal G}}\right)-\frac{1}{2}\left[ F(R,{\cal G})-R\, \frac{\partial F(R,{\cal G})}{\partial  R}-{\cal G} \frac{\partial F(R,{\cal G})}{\partial {\cal G}}\right] =0\,,\nonumber\\ \label{moto11}
\end{eqnarray}

\begin{eqnarray} 
\left[R+6\left( \frac{{\ddot a}}{a}+\frac{{\dot a}^2}{a^2}\right)\right] \frac{\partial^2F(R,{\cal G})}{\partial  R^2}+\left[{\cal G}-24 \left(\frac{{\ddot a}{\dot a}^2}{a^3}\right) \right]\frac{\partial^2 F(R,{\cal G})}{\partial  R \partial {\cal G}}\,=\,0\,,
\label{moto22}
\end{eqnarray}

\begin{eqnarray} 
\left[R+6 \left(\frac{{\ddot a}}{a}+\frac{{\dot a}^2}{a^2}\right)\right]\frac{\partial^2 F(R,{\cal G})}{\partial  R \partial {\cal G}}     +\left[{\cal G}-24 \left(\frac{{\ddot a}{\dot a}^2}{a^3}\right) \right] \frac{\partial^2F(R,{\cal G})}{\partial  {\cal G}^2}=0\,.
\label{moto33}
\end{eqnarray}
It is worth noticing that  Eqs. (\ref{moto22}) and (\ref{moto33})  show a symmetry in the variables $R$ and ${\cal G}$.
Finally the  energy condition  (\ref{energy}),  corresponding to the $00$-Einstein equation, gives

\begin{eqnarray} 
&&\left(\frac{\dot a}{a}\right)^2  \frac{\partial F(R,{\cal G})}{\partial R}+\left(\frac{\dot a}{a}\right) \frac{d}{dt}\left( \frac{\partial F(R,{\cal G})}{\partial R} \right)-4\left(\frac{\dot a}{a}\right)^3 \frac{d}{dt}\left(\frac{\partial F(R,{\cal G})}{\partial {\cal G}}\right) -
 \frac{1}{6} \left[ F(R,{\cal G})-R\, \frac{\partial F(R,{\cal G})}{\partial  R}-{\cal G} \frac{\partial F(R,{\cal G})}{\partial {\cal G}}\right]  \, =\,0\,.\nonumber\\
\label{energy1}
\end{eqnarray}
We can see that considering ${\cal G}$, $R$ and $a$ as  variables,  we have,
for consistency, that $R$ and ${\cal G}$
coincides with the definitions of the Ricci scalar and Gauss-Bonnet invariant in the Friedman-Robertson-Walker
metric, respectively. Geometrically, these are  Euler's constraints of the
dynamics. In the next Sections, we will show how solutions of the system (\ref{moto11}-\ref{energy1}) can be achieved by asking for the existence of Noether symmetries. On the other hand, the existence of the Noether symmetries guarantees the reduction of dynamics allowing the solution of the system.

\section{Noether symmetries in Gauss-Bonnet cosmology}
\label{quattro}

The existence of Noether symmetries allows to select constants of motion so that the
dynamics results simplified. Often such a dynamics is exactly solvable by a straightforward
change of variables where a cyclic ones are determined \cite{safe}.
A Noether symmetry for the Lagrangian (\ref{PointLagra}) exists if the condition 

\begin{eqnarray} 
L_X {\cal L}\,=\,0 \qquad \rightarrow \qquad X{\cal L}\,=\,0\,,
\label{LX}
\end{eqnarray}
holds. Here $L_X$ is the Lie derivative with respect to the Noether vector $X$. Eq.(\ref{LX}) is
nothing else but the contraction of the Noether vector $X$, defined on the tangent space ${\cal TQ}=\{q_i,{\dot q}_i\}=\{a,\dot{a}, R, \dot{R}, {\cal G},{\dot {\cal G}}\}$ of the Lagrangian ${\cal L}={\cal L}(q_i,{\dot q}_i)={\cal L}(a,{\dot a}, R,{\dot R},{\cal G},{\dot {\cal G}})$, with the Cartan one-form, generically defined as

\begin{eqnarray} 
\theta_{\cal L} \equiv \frac{\partial {\cal L}}{\partial {\dot q}_i}dq^i\,.
\end{eqnarray}
Condition (\ref{LX}) gives 

\begin{eqnarray} 
i_X \theta_{\cal L} = \Sigma_0\,,
\end{eqnarray}
where $i_X$ is the inner derivative and $\Sigma_0$ is the conserved quantity \cite{porco,35,36}. In
other words, the existence of the symmetry is connected to the existence of a vector field

\begin{eqnarray} 
X= \alpha^i (q)\frac{\partial}{\partial q^i}+\frac{d\alpha^i(q)}{dt}\frac{\partial}{\partial {\dot q}^i}\,,
\end{eqnarray}
where at least one of the components  $\alpha^i(q)$ have to be different from zero to generate a symmetry. In our case,
the generator of symmetry is 

\begin{eqnarray} 
X=\alpha \frac{\partial}{\partial a}+ \beta\frac{\partial}{\partial R}+\gamma  \frac{\partial}{\partial {\cal G}}+{\dot \alpha} \frac{\partial}{\partial  \dot a}+  {\dot \beta}\frac{\partial}{\partial \dot R}+{\dot \gamma}  \frac{\partial}{\partial\dot {\cal G}}\,.
\label{ourX}
\end{eqnarray}

The functions $\alpha, \beta, \gamma$ depend on the variables $a, R, {\cal G}$ and then
\begin{eqnarray} 
{\dot \alpha}\,=\, \frac{\partial \alpha}{\partial a}{\dot a}+\frac{\partial \alpha}{\partial R}{\dot R}+\frac{\partial \alpha}{\partial {\cal G}}{\dot {\cal G}}\,,\quad
{\dot \beta}\,=\, \frac{\partial \beta}{\partial a}{\dot a}+\frac{\partial \beta}{\partial R}{\dot R}+\frac{\partial \beta}{\partial {\cal G}}{\dot {\cal G}}\,,\quad
{\dot \gamma}\,=\, \frac{\partial \gamma}{\partial a}{\dot a}+\frac{\partial \gamma}{\partial R}{\dot R}+\frac{\partial \gamma}{\partial {\cal G}}{\dot {\cal G}}\,.
\end{eqnarray}

As stated above, a  Noether symmetry exists if at
least one of them is different from zero.  Their analytic forms can be found by making explicit  Eq. (\ref{LX}), which corresponds to a set of partial differential equations given by
equating to zero the terms in ${\dot a}^2, {\dot R}^2, {\dot {\cal G}}^2$, ${\dot a}{\dot R}$, ${\dot a}{\dot{\cal G}}$, ${\dot R}{\dot {\cal G }}$  and so on. In our specific case, we get a system of thirteen partial differential
equations

\begin{equation}
\label{system}
\left\{\begin{array}{ll}
\alpha \frac{\partial F(R,{\cal G})}{\partial  R}+\beta a \frac{\partial^2 F(R,{\cal G})}{\partial  R^2}+\gamma a \frac{\partial F(R,{\cal G})}{\partial  R \partial {\cal G}}+2 a \left(\frac{\partial \alpha}{\partial a}\right) \frac{\partial F(R,{\cal G})}{\partial  R}+ a^2  \left(\frac{\partial \beta}{\partial a}\right)\frac{\partial^2 F(R,{\cal G})}{\partial  R^2}+a^2\left(\frac{\partial\gamma}{\partial}\right)\frac{\partial^2 F(R,{\cal G})}{\partial  R \partial{\cal G}}\,=\,0\\
 \\
2\alpha \frac{\partial^2 F(R,{\cal G})}{\partial  R^2}+\beta a \frac{\partial^3 F(R,{\cal G})}{\partial  R^3}+\gamma a  \frac{\partial^3 F(R,{\cal G})}{\partial  R^2 \partial {\cal G}}+a \left(\frac{\partial \alpha}{\partial a}\right) \frac{\partial^2 F(R,{\cal G})}{\partial  R^2}+2 \left(\frac{\partial \alpha}{\partial R}\right) \frac{\partial F(R,{\cal G})}{\partial  R}+a  \left(\frac{\partial \beta}{\partial R}\right)\frac{\partial^2 F(R,{\cal G})}{\partial  R^2}+a\left(\frac{\partial \gamma}{\partial R}\right)\frac{\partial^2 F(R,{\cal G})}{\partial  R \partial{\cal G}}
\,=\,0\\
\\
\beta\, \frac{\partial^3 F(R,{\cal G})}{\partial {\cal G}^2 \partial  R}+\gamma\,\frac{\partial^3 F(R,{\cal G})}{\partial {\cal G}^3}+3  \left(\frac{\partial \alpha}{\partial a}\right)\frac{\partial^2 F(R,{\cal G})}{\partial {\cal G}^2}+ \left(\frac{\partial \beta}{\partial {\cal G}} \right)\frac{\partial^2 F(R,{\cal G})}{\partial  R \partial{\cal G}}+ \left(\frac{\partial \gamma}{\partial a}\right)\frac{\partial^2 F(R,{\cal G})}{\partial {\cal G}^2}\,=\,0\\
\\
 \left(\frac{\partial \alpha}{\partial R}\right) \frac{\partial^2 F(R,{\cal G})}{\partial  R^2}\,=\,0\\
 \\
  \left(\frac{\partial \alpha}{\partial R}\right) \frac{\partial^2 F(R,{\cal G})}{\partial  {\cal G}^2}+\left(\frac{\partial\alpha}{\partial{\cal G}}\right)\frac{\partial^2 F(R,{\cal G})}{\partial  R \partial{\cal G}}\,=\,0\\
\\
\left(\frac{\partial \alpha}{\partial R}\right)\frac{\partial^2 F(R,{\cal G})}{\partial  R \partial{\cal G}}    +\left(\frac{\partial \alpha}{\partial {\cal G} }\right) \frac{\partial^2 F(R,{\cal G})}{\partial R^2}\,=\,0\\
\\
2 \alpha \frac{\partial ^2 F(R,{\cal G})}{\partial R \partial {\cal G}}+ \beta a  \frac{\partial^3 F(R,{\cal G})}{\partial {\cal G} \partial  R^2} +\gamma a \frac{\partial^3 F(R,{\cal G})}{\partial {\cal G}^2 \partial  R}+a\left(\frac{\partial\alpha}{\partial a}\right)\frac{\partial^2 F(R,{\cal G})}{\partial  R \partial{\cal G}}+ 
+2\left(\frac{\partial \alpha}{\partial {\cal G} }\right)  \frac{\partial F(R,{\cal G})}{\partial R}
+a \left(\frac{\partial \beta}{\partial {\cal G} }\right) \frac{\partial^2 F(R,{\cal G})}{\partial R^2}+a \left(\frac{\partial \gamma}{\partial {\cal G}}\right)\frac{\partial^2 F(R,{\cal G})}{\partial R \partial {\cal G}}\,=\,0\\
\\
\left(\frac{\partial\beta}{\partial a}\right)\frac{\partial^2 F(R,{\cal G})}{\partial R \partial {\cal G}}\ +\left(\frac{\partial \gamma}{\partial a}\right)\frac{\partial^2 F(R,{\cal G})}{\partial  {\cal G}^2}\,=\,0\\
\\
\beta  \frac{\partial^3 F(R,{\cal G})}{\partial {\cal G} \partial  R^2}+\gamma  \frac{\partial^3 F(R,{\cal G})}{\partial {\cal G}^2 \partial  R}+\left(\frac{\partial\beta}{\partial R}\right)\frac{\partial^2 F(R,{\cal G})}{\partial R \partial {\cal G}}+   \left(\frac{\partial \gamma}{\partial R}\right)\frac{\partial^2 F(R,{\cal G})}{\partial  {\cal G}^2}+3 \left(\frac{\partial \alpha}{\partial a}\right)\frac{\partial^2 F(R,{\cal G})}{\partial R \partial {\cal G}}\,=\,0\,\\
\\
\left(\frac{\partial \alpha}{\partial {\cal G}}\right)\frac{\partial^2 F(R,{\cal G})}{\partial  {\cal G}^2}\,=\,0\,\\
\\
\left(\frac{\partial \alpha}{\partial R}\right)\frac{\partial^2 F(R,{\cal G})}{\partial R \partial  {\cal G}}\,=\,0\\
\\
\left(\frac{\partial \alpha}{\partial {\cal G}}\right)\frac{\partial^2 F(R,{\cal G})}{\partial R \partial  {\cal G}}\,=\,0\\
\\
3\alpha \left[ F(R,{\cal G})-R\, \frac{\partial F(R,{\cal G})}{\partial  R}-{\cal G} \frac{\partial F(R,{\cal G})}{\partial {\cal G}}\right]-\beta a \left[ R\, \frac{\partial^2 F(R,{\cal G})}{\partial  R^2}+{\cal G}\, \frac{\partial^2 F(R,{\cal G})}{\partial {\cal G}\partial  R}\right]-\gamma a \left[ {\cal G}\, \frac{\partial^2 F(R,{\cal G})}{\partial  {\cal G}^2}+R\, \frac{\partial^2 F(R,{\cal G})}{\partial  R\partial {\cal G}}\right]\,=\,0\\
\end{array}\right.\end{equation}

The system (\ref{system}) is overdetermined and, if solvable, enables one to assign $\alpha,\beta,\gamma$ and $F(R,{\cal G})$.
The analytic form of $F(R,{\cal G})$ can be fixed by imposing, in the last equation of system (\ref{system}), the conditions

\begin{equation}\label{siste13}
\left\{\begin{array}{ll}
 F(R,{\cal G})-R\, \frac{\partial F(R,{\cal G})}{\partial  R}-{\cal G} \frac{\partial F(R,{\cal G})}{\partial {\cal G}}=0\\
\\
R\, \frac{\partial^2 F(R,{\cal G})}{\partial  R^2}+{\cal G}\, \frac{\partial^2 F(R,{\cal G})}{\partial {\cal G}\partial  R}=0\\
\\
 {\cal G}\, \frac{\partial^2 F(R,{\cal G})}{\partial  {\cal G}^2}+R\, \frac{\partial^2 F(R,{\cal G})}{\partial  R\partial {\cal G}}=0\\
\end{array}\right.\end{equation}
where the second and third equations  are  symmetric. However, it is clear that this is nothing else but an arbitrary choice since more general conditions are possible.  In particular, we can choose the functional forms:

\begin{eqnarray} 
F(R,{\cal G}) = f(R)+ f({\cal G})\,, \qquad F(R,{\cal G}) = f(R) f({\cal G})\,,
\end{eqnarray} 
from which it is easy to prove that the functional forms compatible with the system (\ref{siste13}) are:

\begin{eqnarray} 
F(R,{\cal G})= F_0 R+ F_1 G\,,\qquad 
F(R,{\cal G})= F_0 R^n {\cal G}^{1-n}\,.
\end{eqnarray}
This allows to work out cosmological models compatible with the Noether symmetries.

\section{Examples of exact cosmological solutions}
\label{cinque}

\subsection{The case  $F(R,{\cal G})=F_0R+F_1{\cal G}$}
\label{cinque1}
The Noether symmetry related to the functional form of $F(R,{\cal G})=F_0R+F_1{\cal G}$ can be achieved considering the system (\ref{system}).
The first equation gives
\begin{eqnarray} 
\alpha+2a\frac{d\alpha}{da}=0\,,
\label{si1}
\end{eqnarray}
while the others, a part the last one,  are identically zero.
Immediately, the Noether symmetry is given by

\begin{eqnarray} 
 \alpha=\frac{\alpha_0}{\sqrt{a}}\,, \qquad \beta=0, \qquad \gamma=0\,.
\end{eqnarray}
This theory is nothing else but General Relativity  where the Gauss-Bonnet term  is vanishing in 4 dimension.
In fact, we have
\begin{eqnarray} 
\left(\frac{\dot a}{a}\right)^2+2 \frac{\ddot a}{a}&=&0\,,\\
 R+ 6 \left[\left(\frac{\dot a}{a}\right)^2+ \frac{\ddot a}{a}\right]&=&0\,,\\
{\cal G}-24\left(\frac{{\ddot a}{\dot a}^2}{a^3}\right)&=&0\,,\\
\left(\frac{\dot a}{a}\right)^2&=&0\,,
\end{eqnarray}
that is the Minkowski spacetime is recovered in vacuum while standard Friedman solutions are recovered considering standard perfect fluid matter.

\subsection{The case $F(R,{\cal G})= F_0 R^n {\cal G}^{1-n}\,,$}

This is a more interesting case where the Noether symmetry rules the relation between the two invariant scalars $R$ and ${\cal G}$.
Let us choose the simplest non-trivial case $n = 2$. The functional form of $F(R,{\cal G})$ becomes  ${\displaystyle F(R,{\cal G})= F_0 \frac{R^2}{ {\cal G}}}$. Then the point-like Lagrangian is
\begin{eqnarray} 
{\cal L}=\frac{4\, F_0\, {\dot a}}{\cal G}\left[3\, a\, {\dot a}\, R+3\,a\,{\dot R}-3\, a^2 \,{\dot {\cal G}} \frac{R}{\cal G}+4\, {\dot a}^2 \,{\dot R}\frac{R}{\cal G}-4\, {\dot a}^2\,{\dot {\cal G}}\left(\frac{R}{\cal G}\right)^2\right]\,.
\label{lagrangian_n2}
\end{eqnarray}
To solve the system, we can choose the variable $ {\displaystyle \frac{R}{\cal G}= \zeta}$
and then the Lagrangian (\ref{lagrangian_n2}) is transformed into
\begin{eqnarray} 
{\cal L}= 12\, a\,{\dot a}^2 F_0 \zeta+12 a^2 {\dot a} F_0 {\dot \zeta}+16 {\dot a}^3F_0  \zeta {\dot \zeta}\,.
\label{lagrangian_zeta}
\end{eqnarray}
Clearly we have reduced the dynamics  assuming that $\zeta$  depends on $R$ and ${\cal G}$.
The Euler-Lagrange equations are 
\begin{eqnarray}
&& 6\left[\frac{\ddot a}{a}+\left(\frac{\dot a}{a}\right)^2\right]+24\left(\frac{{\ddot a} {\dot a}^2}{a^3}\right)\zeta=0\,,
\label{motoz1}\\
&& \left[2\frac{\ddot a}{a}+\left(\frac{\dot a}{a}\right)^2\right]\zeta+2\frac{\dot a}{a} {\dot \zeta}+{\ddot\zeta}+8\left( \frac{{\dot a}{\ddot a}}{a^2}\right)\zeta{\dot \zeta}+4\left(\frac{\dot a}{a}\right)^2{\dot \zeta}^2+4 \left(\frac{\dot a}{a}\right)^2\zeta {\ddot\zeta}=0\,.
\label{motoz2}
\end{eqnarray} 
and the energy condition 
\begin{eqnarray}
 \left(\frac{\dot a}{a}\right)^2+\left(\frac{\dot a}{a}\right)\left(\frac{\dot \zeta}{\zeta}\right)+4\left(\frac{\dot a}{a}\right)^3 {\dot \zeta}=0\,.
\label{energyz}
\end{eqnarray} 
Eq.(\ref{motoz1}) is a consistency condition which is immediately verified as soon as definitions (\ref{consistency}) are replaced.
Power law solutions are  found, being
\begin{eqnarray}
 a(t)=a_0t^s\,,  \qquad  \zeta=\zeta_0t^2\,, \quad \mbox{with}\quad  s=3\,,
 \end{eqnarray}
being $\zeta_0=-5/72$.  

Another solution  is  
\begin{eqnarray}
a(t)=a_0 \exp(\Lambda t)\,,\qquad
\ln \left(\frac{\zeta}{\zeta_0}\right)+4\Lambda^2\left(\zeta - \zeta_0\right)=-\Lambda t\,,
\end{eqnarray}
where $\Lambda$ is a constant. It is clear that ${\cal G}$ always evolve as $R^2$ as it must be.

Other cases can be worked out. Here we have just given some examples where the existence of the symmetry allows a suitable reduction of the dynamical system and a full control of the problem based on first principles.

\section{Conclusions}
\label{sei}
The above discussion points out that the existence of Noether symmetries is capable of selecting suitable $F(R,{\cal G})$ models on the basis of first principles and then to integrate dynamics by the identification of suitable cyclic variables.  A consequent reduction process  allows to achieve exact solutions. A main role is played by the  Lagrange multipliers that make the point-like Lagrangian canonical. This key feature is extremely important since allows to deal with invariants like $R$ and ${\cal G}$ under the same standard of auxiliary effective scalar fields. In fact, from a general point of view, $R$ and ${\cal G}$ are functions of $\{a,\dot{a},\ddot{a}\}$ so dynamics is not a priori canonical. The Lagrange multipliers permit to overcome this difficulties and to deal with the further degrees of freedom related to $F(R,{\cal G})$ \cite{makarenko}. 

However, it is worth stressing that symmetries are not only a mathematical tool to solve dynamics. As discussed in \cite{hamilton}, their existence allows to select physically observable universes  and, specifically, to select analytical models related to observations. In particular,  it is possible to classify dark energy behaviors related to Noether symmetries and then discriminate between physical and unphysical solutions (see  \cite{greci,basilakos,basilakos1,basilakos2}). These topics will be the argument of a forthcoming, detailed study.

\section*{Acknowledgments}
 SC and MDL  are supported by INFN Sez. di Napoli ({\it Iniziative
Specifiche} QGSKY and TEONGRAV). SDO  is supported   by MINECO (Spain), FIS2010-15640 and by MES project
TSPU-139 (Russia).


\end{document}